%Paper: hep-th/9301006
%From: "Kresimir Demeterfi" <kresimir@puhep1.Princeton.EDU>
%Date: Tue, 5 Jan 93 13:19:49 EST

% PHYZZX file begines here

\input phyzzx

\sequentialequations

                        %Real Nucl. Phys. format
\overfullrule=0pt
\catcode`\@=11
\def\mm{matrix model}
\def\ns{non-singlet}

\def\NP{{\it Nucl. Phys.\ }}

\def\PL{{\it Phys. Lett.\ }}
\def\PR{{\it Phys. Rev.\ }}
\def\PRL{{\it Phys. Rev. Lett.\ }}
\def\CMP{{\it Comm. Math. Phys.\ }}

\def\Mod{{\it Mod. Phys. Lett.\ }}

\def\td{two-dimensional}
\def\lc{light-cone}
\def\KT{Kosterlitz-Thouless}
\def\eqaligntwo#1{\null\,\vcenter{\openup\jot\m@th
\ialign{\strut\hfil
$\displaystyle{##}$&$\displaystyle{{}##}$&$\displaystyle{{}##}$\hfil
\crcr#1\crcr}}\,}
\catcode`\@=12

\REF\GM{D.~J.~Gross and A.~A.~Migdal, \PRL {\bf 64} (1990) 717;
M. Douglas and S.~Shenker, \NP {\bf B335} (1990) 635;
E.~Br\'ezin and V.~Kazakov, \PL {\bf 236B} (1990) 144.}
\REF\Doug{ M. Douglas, \PL {\bf 238B} (1990) 176.}
\REF\GMil{D. J. Gross and N. Miljkovi\'c, \PL
 {\bf 238B} (1990) 217; E. Br\'ezin, V. Kazakov, and Al. B. Zamolodchikov,
\NP {\bf B338} (1990) 673; P. Ginsparg and J. Zinn-Justin,
\PL{\bf  240B} (1990) 333; G. Parisi, \PL {\bf 238B} (1990) 209.}
\REF\GKl{D. J. Gross and I. R. Klebanov, \NP {\bf B344} (1990) 475.}
\REF\DJ{S. Das and A. Jevicki, \Mod {\bf A5} (1990) 1639;
J. Polchinski, \NP {\bf B346} (1990) 253. }
\REF\barbon{L. Alvarez-Gaum\'e, J. Barb\'on and \v C. Crnkovi\'c,
CERN preprint CERN-TH-6600/92 and FTUAM-9223, July 1992.}
\REF\hikami{E. Br\'ezin and S. Hikami, \PL {\bf 283B} (1992) 203;
LPTENS preprint 92-31, September 1992.  }
\REF\dur{B. Durhuus, J. Fr\" ohlich and T. Jonsson,
\NP {\bf B240} (1984) 453. }
\REF\kutseib{N. Seiberg, {\it Prog. Theor. Phys. Suppl.} {\bf 102}
(1990) 319;
D. Kutasov and N. Seiberg, \NP {\bf B358} (1991) 600.}
\REF\DK{S. Dalley and I. R. Klebanov, Princeton preprint
PUPT-1333, July 1992, to appear in Phys. Lett. B.}
\REF\KS{I. R. Klebanov and L. Susskind, \NP {\bf B309} (1988) 175.}
\REF\I{For a review, see I. Klebanov, Princeton preprint PUPT-1271,
in {\sl String Theory and Quantum Gravity `91}, World Scientific, 1992.}
\REF\Vort{D. J. Gross and I. R. Klebanov,
\NP {\bf B354} (1991) 459.}
\REF\bk{D. Boulatov and V. Kazakov, LPTENS preprint 91-24, August 1991.}
\REF\thooft{J. Koplik, A. Neveu and S. Nussinov, \NP {\bf B123}
(1977) 109;
G. 't Hooft, \CMP {\bf 86} (1982) 449; {\bf 88} (1983) 1.}
\REF\chang{See, for instance, S. J. Chang, R. G. Root and T. M. Yan
\PR {\bf D7} (1973) 1133.}
\REF\Thorn{C. Thorn, \PR {\bf D17} (1978) 1073.}
\REF\Brod{H.-C. Pauli and S. Brodsky, \PR {\bf D32} (1985) 1993, 2001;
K. Hornbostel, S. Brodsky and H.-C. Pauli, \PR {\bf D41} (1990) 3814;
for a good review, see K. Hornbostel, Ph. D. thesis, SLAC report
No. 333 (1988).}
\REF\Gauged{S. Dalley and I. R. Klebanov, Princeton preprint
PUPT-1342, September 1992, to appear in Phys. Rev. D.}

\def\eqaligntwo#1{\null\,\vcenter{\openup\jot\m@th
\ialign{\strut\hfil
$\displaystyle{##}$&$\displaystyle{{}##}$&$\displaystyle{{}##}$\hfil
\crcr#1\crcr}}\,}
\catcode`\@=12

\def\qg{quantum gravity}

\def\half{{1\over 2}}
\def\d{\dagger}

\nopagenumbers

\line{\hfil PUPT-1370}
\line{\hfil December 1992}
\title{Light-Cone Approach  to Random Surfaces Embedded in
Two Dimensions\foot{Based on the talk delivered by I. R. Klebanov at the
7th Nishinomiya--Yukawa Memorial Symposium ``Quantum Gravity'',
November 1992.} }
\author {Kre\v simir Demeterfi\foot{{\rm On leave of absence from
the Rudjer Bo\v skovi\'c Institute, Zagreb, Croatia}}
 and Igor R. Klebanov }
\JHL
\abstract
We review the recently proposed \lc\ quantization of the
matrix model which is expected to have
a critical point describing
2-d quantum gravity coupled to $c=2$ matter.
In the $N\to\infty$ limit, we derive a linear
Schroedinger equation for the free string spectrum. Numerical study of
this equation suggests that the spectrum is tachyonic, and that
the string tension diverges at the critical point.
\endpage

\pagenumbers
\vsize=8.9in
\hsize=6.5in
\centerline{\caps 1. Introduction}
\bigskip

Recently there has been  considerable renewed interest in
large-$N$ \mm s. The one-dimensional hermitian matrix chain models
have been solved in the double-scaling limit and identified with
the $c<1$ minimal models coupled to \td\ \qg [\GM, \Doug].
In the same fashion
the hermitian matrix quantum mechanics, which describes the $c=1$ theory,
has been solved both for non-compact [\GMil] and for circular
target space [\GKl]. The $c=1$
model can be interpreted in terms of  $D=2$ string theory [\DJ], where
the role of the extra dimension is played by the conformal factor
of the world sheet \qg. One should keep in mind, however, that the models
with $c>1$ are of much greater interest because they are expected
to correspond to string theories in $D>2$. These theories should
have a much richer structure than in $D=2$ because strings
can exhibit transverse oscillations. In spite of some interesting new
developments [\barbon, \hikami],
there is little solid information
about the $c>1$ theories.

{}From the existing
analytical and Monte Carlo studies there is some evidence that the
discretized $c>1$ models are in the branched polymer phase [\dur]
and do not lead to
acceptable bosonic string theories. On the other hand, the continuum
approach indicates that these models are tachyonic [\kutseib]. It is often
suggested that the branch polymer behavior is indicative of the presence
of the tachyon. However, some of the discretized models should
not be tachyonic due to reflection positivity [\dur].
We believe that
a deeper understanding of the $c>1$ models is needed because, even though
the simplest such theories are probably unacceptable, there may be
modifications that lead to new interesting string models. With this idea
in mind, Dalley and one of the authors [\DK] have recently attempted a new
approach to the theory of random surfaces embedded in two dimensions
($c=2$). The corresponding matrix model is a two-dimensional
super-renormalizable scalar field theory which can be quantized using
the light-cone coordinates. The $N\to\infty$ limit naturally leads
to a free string \lc\ Schroedinger equation [\KS]. In these notes we will
review the light-cone quantization of this model, and will report on
new numerical studies of the free string spectrum which suggest
a different kind of behavior than what was anticipated in ref. \DK.

\endpage

\bigskip
\centerline{\caps 2. Random Surfaces Embedded in One and Two Dimensions}
\bigskip

Before proceeding to the $c=2$ model, we will briefly review some of the
lessons learned from the exactly soluble $c=1$ matrix model for the
purpose of comparison. The discretized approach to 2-d quantum gravity
coupled to $c=1$ matter [\I] is generated by the euclidean matrix
quantum mechanics with the action
$$\int dx \Tr \biggl (\half
\left({\partial M\over \partial x}\right)^2 + {1\over 2\alpha'} M^2 -
{\lambda\over 3\sqrt N} M^3 \biggl )\,\,,
\eqn\qm$$
where $M(x)$ is an $N\times N$ hermitian matrix.
The connection of this \mm\ with triangulated random surfaces follows,
as usual, after identifying the Feynman graphs with the graphs
dual to triangulations. The lattice link factor is the one-dimensional
scalar propagator,
$$ G(x_i, x_j)={\sqrt{\alpha'}\over 2}
\exp \left(-|x_i-x_j|/\sqrt{\alpha'}\right)
\eqn\link$$
so that the parameter $\alpha'$ sets the
scale in the embedding dimension. The model \qm\ possesses a
global $SU(N)$ symmetry under $M\to \Omega^\dagger M\Omega$.
It is well-known that the $SU(N)$
singlet spectrum is described by $N$ free fermions moving in the
cubic potential, with the Planck constant $\sim 1/N$. In the leading WKB
approximation the ground state energy is $\sim N^2$,
which corresponds to the sum over surfaces of spherical topology.
This sum is singular as $\Delta\sim\lambda_c-\lambda\to 0$, which is the
value of the coupling where the Fermi level reaches the local maximum
of the cubic potential,
$$ {\cal Z}_0={1\over \sqrt {\alpha'}} {N^2\Delta^2\over \ln\Delta}\,\,.
$$
The geometrical meaning of this singularity
is that the Feynman graphs with very large numbers of vertices
are becoming important. Near this singularity the universal continuum
limit of the 2-d quantum gravity coupled to $c=1$ matter
can be defined. In this limit,
the gaps in the low-lying spectrum go to zero as $1/|\ln\Delta|$.
The underlying dispersion relation [\DJ]
$$ \alpha' E^2-p_\phi^2 =0$$
is characteristic of a massless two-dimensional field theory. The hidden
$\phi$-dimension derives from the fluctuating conformal factor of the
world sheet. It is worth emphasizing that the string scale $\alpha'$
does not undergo any singular renormalization in the $\Delta\to 0$
limit and essentially corresponds to the ``bare'' string scale defined
by the link factor \link.

An important fact about the $c=1$ matrix model is the decoupling of the
$SU(N)$ non-singlet states, whose energies diverge as $|\ln\Delta|$.
As a result, only the singlet states are important for random surfaces
embedded in a non-compact dimension, or in a circle of radius $R>R_c$
[\GKl].
The decoupling of the non-singlet states corresponds physically to
the confinement of Kosterlitz-Thouless vortices [\Vort, \bk].

In ref. \DK\ a generalization of the matrix model \qm\ to two dimensions was
studied,
$$ S_E=\int d^2 x \Tr \left (\half (\partial_\alpha M)^2+\half \mu M^2-
{1\over 3\sqrt{N}}\lambda M^3 \right )\ ,
\eqn\action$$
where $M(x^0, x^1)$ is an $N\times N$ hermitian matrix field.
Just as for $c=1$, the Feynman graphs of this theory are identified
with the graphs
dual to triangulations. The only change is that
now the lattice link factor is the two-dimensional
scalar propagator,
$$ G(\vec x_i, \vec x_j)= \int {d^2 p \over (2\pi)^2 }{{\rm e}^{i
\vec p \cdot (\vec x_i -\vec x_j) }
\over p^2+\mu} = {1\over 2\pi} K_0 \bigl(\sqrt\mu~ |\vec x_i-\vec x_j|
\bigr)
\ ,\eqn\eq$$
where $K_0$ is a modified Bessel function.
Thus, at the leading
order in $N$, we obtain
a sum over the planar triangulated random surfaces
embedded in two dimensions.\foot{Note that, as for $c=1$, the
coordinates are specified at the centers of the triangles.}
The logarithmic divergence of $G$ at small separations is very mild.
If the tadpole graphs are discarded, as they should be because they
do not correspond to good triangulations, then each separate
Feynman graph is finite.
This is similar to what we find in the matrix models for $c\leq 1$.
For these types of theories there are general arguments [\thooft] indicating
that, for sufficiently small $\lambda/\mu$, the sum over planar graphs is
finite.\foot{This is essentially because the number of planar graphs
grows only exponentially with the number of vertices, and one can put
exponential bounds on the value of each graph.}
Therefore, we may look for singularities in various
physical quantities
at some critical value of the dimensionless parameter
$\lambda/\mu$. The crucial question is whether, as for $c=1$,
a sensible string theory can be defined near this critical coupling.
Although the answer is almost certainly negative, it is interesting
to study the nature of the critical point. In these notes we will
describe our recent numerical results which suggest that the spectrum
of the theory is tachyonic, and that the string tension actually
diverges at the critical point. We will also speculate on how to find
a cure for this undesirable behavior.
\bigskip
\centerline{\caps 3. Light-Cone Quantization}
\bigskip

Although the $c=2$ \mm\ of eq.
\action\ is certainly more complex than the matrix quantum mechanics,
we should still be able to take advantage of the simplifications
that distinguish \td\ field theories. The method to pursue this
that seems particularly convenient, is to continue $x^0\to ix^0$
and to carry out \lc\ quantization [\chang] of the resulting
$(1+1)$-dimensional field theory,
treating $x^{+}=(x^0+x^1)/\sqrt 2$
as the time and $x^{-}=(x^0-x^1)/\sqrt 2$ as the spatial
variable. From the Minkowski signature action,
$$ S=\int dx^+ dx^- \Tr \left (\partial_+ M\partial_-M-\half \mu M^2+
{1\over 3\sqrt{N}}\lambda M^3 \right )\ ,
\eqn\miaction$$
we derive the light-cone components of the total momentum
$$ \eqalign{&P^{+}(x^{+}) = \int dx^{-} \Tr\, (\partial_{-}M)^{2}\ ,\cr
&P^{-}(x^{+}) = \int dx^{-} \Tr\, (\half \mu M^2 - {\lambda \over 3\sqrt{N}}
M^3 )\ .\cr }\eqn\pminus$$
Upon quantization, the commutation relations are imposed at equal $x^{+}$:
$$ [M_{ij}(x^{-}),\partial_{-}M_{kl}(\tilde{x}^{-})] = \half i\delta (x^{-} -
\tilde{x}^{-})\delta_{il} \delta_{jk}\ .\eqn\rel$$
These commutation relations
differ from the canonical by the extra factor of
$\half$, which follows from the appropriate constrained quantization.
This extra factor ensures, for instance, that
$[P^+, M]=\partial_-M$.
Expanding in Fourier modes\foot{The symbol $\dagger$ is always understood
to have purely quantum meaning and does not act on indices.},
$$M_{ij}={1 \over \sqrt{2\pi}} \int_{0}^{\infty} {dk^{+} \over \sqrt{2k^{+}}}
\biggl (a_{ij}(k^{+}){\rm e}^{-ik^{+}x^{-}} + a_{ji}^{\d}(k^{+})
{\rm e}^{ik^{+}x^{-}} \biggr)\eqn\Four$$
we find the conventional oscillator algebra
$$[a_{ij}(k^{+}),a_{lk}^{\d}(\tilde{k}^{+})] = \delta(k^{+} - \tilde{k}^{+})
\delta_{il}\delta_{jk}\ .\eqn\modeccr$$
After substituting eq. \Four\ into eq. \pminus\ and normal ordering, we obtain
$$\eqalign{
&P^+ = \int_{0}^{\infty} dk^+ k^+ a_{ij}^{\d}(k^+)a_{ij}(k^+)\ ,\cr
&P^{-} =
\half \mu \int_{0}^{\infty} {dk^{+}\over k^{+}} a_{ij}^{\d}(k^{+})
a_{ij}(k^{+}) -{\lambda \over 4\sqrt{N\pi}}\cr
&\times\int_{0}^{\infty}
{dk_{1}^{+}dk_{2}^{+}\over \sqrt{k_{1}^{+}k_{2}^{+}(k_{1}^{+} + k_{2}^{+})}}
\left\{a_{ij}^{\d}(k_1^+ +k_2^+)a_{ik}(k_2^+)a_{kj}(k_1^+) +
a_{ik}^{\d}(k_1^+)a_{kj}^{\d}(k_2^+)a_{ij}(k_1^+ +k_{2}^+)\right\}
\cr }\eqn\nominus$$
(repeated indices are summed over). The normal
ordering is equivalent to removing the tadpole graphs in the Lagrangian
approach. $P^+$ commutes with $P^-$ so
that they can be diagonalized simultaneously.
As usual in light-cone quantization, the vacuum $|0 \rangle$,
which is defined by
$$ a_{ij} (k^+) |0 \rangle =0\ ,\eqn\eq$$
is an eigenstate of the fully interacting \lc\ Hamiltonian
$P^{-}$ with eigenvalue zero.
This is related to the fact that the light-cone momentum
$k^+$ is positive for all quanta.
We will find that, for sufficiently large $\lambda/\mu$,
$|0 \rangle$ is not the ground state.

Our goal is to study the ground state and the low-lying excitations.
Due to the global $SU(N)$ symmetry, $M\to \Omega^{\dagger} M\Omega$,
the eigenstates can be classified according to their transformation
properties. We expect the low-lying states to transform as singlets
under the $SU(N)$. A general singlet state, carrying light-cone momentum
$P^+$, can be written as
$$\eqalign{|\Psi(P^+) \rangle
=&\sum_{b=1}^\infty \int_0^{P^+} dk_1 dk_2\ldots dk_b \,
\delta\left (\sum_{i=1}^b k_i-P^+\right ) \cr & f_b (k_1, k_2, \ldots, k_b)
N^{-b/2} \Tr [a^{\d}(k_1)\ldots a^{\d}(k_b)] |0 \rangle\ , \cr }
\eqn\eq$$
where we have dropped the superscripts $+$ on $k_i$ for brevity.
Due to the cyclic property of the trace, the functions
$f_b$ trivially obey the cyclic symmetry
$$f_b (k_1, k_2, \ldots, k_b)=f_b (k_b, k_1, \ldots, k_{b-1})=
\ldots =f_b (k_2, k_3, \ldots, k_b, k_1)
\eqn\eq$$
{}From the normalization condition
$$\langle\Psi(P'^+) |\Psi(P^+) \rangle = \delta (P^+- P'^+)
$$
we obtain
$$\sum_{b=1}^\infty  b\int_0^{P^+} dk_1 dk_2\ldots dk_b \,
\delta\left (\sum_{i=1}^b k_i-P^+\right )
|f_b (k_1, k_2, \ldots, k_b)|^2 =1 \,\,.
\eqn\norm$$

The singlet states play a special role in this theory: they have no
$SU(N)$ degeneracy factors and can be thought of as closed strings.
Each $a_{ij}^{\d}(k)$ creates a string bit carrying
longitudinal momentum $k$,
and a state of the form
$$\int_0^{P^+} dk_1 dk_2\ldots dk_b \,
\delta\left (\sum_{i=1}^b k_i-P^+\right ) f_b (k_1, k_2, \ldots, k_b)
N^{-b/2} \Tr [a^{\d}(k_1)\ldots a^{\d}(k_b)] |0\ \rangle  \,.
$$
has the interpretation of a closed string which is a bound state of $b$
string bits.
The function $f_b(k_1, k_2, \ldots, k_b)$ superposes
states with different distributions of longitudinal momentum
along the string. If a large-$N$ model is to correspond
to a reasonable string theory, it should be energetically favorable for all the
oscillator indices to be contracted, so that the quanta of the matrix
field are bound into strings. In fact, if there are finite energy states
with free indices, then at sufficiently high energy the composite
strings are unstable with respect to disintegration into bits.
As we will discuss in the conclusion, this disease is likely to occur in
any model with no local gauge symmetry.

For now, we restrict ourselves
to the study of $SU(N)$ singlets, which are the good string states.
When acting on such states, the light-cone Hamiltonian $P^-$ has an
important property:
in the limit $N \to \infty$,
$P^{-}$ takes single closed string states into single closed string states.
One can easily check that the terms that convert one closed string into
two closed strings (two oscillator traces acting on the vacuum) are
of order $1/N$. Thus, as expected, the string coupling constant is
$\sim 1/N$, and sending it to zero
allows us to study the spectrum of free closed string states.
The resulting linear \lc\ Schroedinger equation can be expressed
as a set of integral equations for the functions $f_b$,
$$\eqalign{& P^- f_b (k_1, k_2, \ldots, k_b)= {\mu\over 2}
f_b (k_1, k_2, \ldots, k_b)\sum_{i=1}^b {1\over k_i}\cr
& -{\lambda\over 4\sqrt\pi}\biggl\{
\int_0^{k_1} dk'_1 {f_{b+1} (k'_1, k_1-k'_1, k_2, \ldots, k_b)\over
\sqrt {k_1 k'_1 (k_1-k'_1)}}+
\int_0^{k_2} dk'_2 {f_{b+1} (k_1, k'_2, k_2-k'_2, k_3, \ldots, k_b)\over
\sqrt {k_2 k'_2 (k_2-k'_2)}}+\ldots \cr
&+ {f_{b-1} (k_1+k_2, k_3, \ldots, k_b)\over \sqrt {k_1 k_2 (k_1+k_2)}}+
{f_{b-1} (k_1, k_2+k_3, \ldots, k_b)\over \sqrt {k_2 k_3 (k_2+k_3)}}+
\ldots \biggr\}\,\,.
\cr }\eqn\lightS$$
The terms proportional to $\lambda$ arise either from replacing
two neigboring bits by one or from dividing a bit into two neigboring
bits. The Lorentz invariance of eq. \lightS\ can be made explicit
if we introduce the longitudinal momentum fractions $x_i=k_i/P^+$,
and rewrite the equation in terms of the functions
$$ \tilde f_b (x_1, x_2, \ldots, x_b)= (P^+)^{(b-1)/2}
f_b (x_1 P^+, x_2 P^+, \ldots, x_b P^+) \,\,.
$$
{}From eq. \lightS\ it is not hard to derive
$$\eqalign{& 2P^+P^- \tilde f_b (x_1, x_2, \ldots, x_b)= \mu
\tilde f_b (x_1, x_2, \ldots, x_b)\sum_{i=1}^b {1\over x_i}\cr
& -{\lambda\over 2\sqrt\pi}\biggl\{
\int_0^{x_1} dx_1' {\tilde f_{b+1} (x'_1, x_1-x'_1, x_2, \ldots, x_b)\over
\sqrt {x_1 x'_1 (x_1-x'_1)}}+
\int_0^{x_2} dx_2' {\tilde f_{b+1} (x_1, x'_2, x_2-x'_2, x_3, \ldots, x_b)\over
\sqrt {x_2 x'_2 (x_2-x'_2)}}+\ldots \cr
&+ {\tilde f_{b-1} (x_1+x_2, x_3, \ldots, x_b)\over \sqrt {x_1 x_2 (x_1+x_2)}}+
{\tilde f_{b-1} (x_1, x_2+x_3, \ldots, x_b)\over \sqrt {x_2 x_3 (x_2+x_3)}}+
\ldots \biggr\} \,\, ,
\cr }\eqn\newlightS$$
and from eq. \norm\ we find that the functions $\tilde f_b$
are normalized according to
$$\sum_{b=1}^\infty  b\int_0^1 dx_1 dx_2\ldots dx_b\,
\delta\left (\sum_{i=1}^b x_i-1\right )
|\tilde f_b (x_1, x_2, \ldots, x_b)|^2 =1
\ .\eqn\newnorm$$
$2P^+P^-$ is manifestly Lorentz invariant, and its eigenvalues
are the squared masses of the closed string states.

Eqs. \newlightS\ and \newnorm\ define a rather unusual eigenvalue
problem which involves an infinite number of unknown functions
$\tilde f_b$. Each of these functions depends on a number of real
variables $x_i$ confined to the interval $[0, 1]$, with the constraint
$\sum_{i=1}^b x_i=1$. An important issue is the behavior of the
functions as $x_i\to 0$. For $\mu>0$ the behavior
$$\lim_{x_1\to 0}\tilde f_b (x_1, x_2, \ldots, x_b) \neq 0
$$
is inconsistent with a finite energy eigenstate because the contribution
$\sim \mu$ blows up near $x_1=0$ faster than the other terms.
Guided by this consideration, we will assume that the functions
$\tilde f_b$ approach zero as the variables approach the end-points
of the interval.

The system of equations \newlightS\ is rather difficult, and we do
not know its exact solution. For that reason, we will attempt to
estimate the spectrum numerically, after introducing a cut-off.
In \lc\ quantization, a particularly convenient method is to replace
the continuous momentum fractions $x$ by a discrete set $n/K$,
where the positive integer $K$ is sent to infinity as the cut-off
is removed [\Thorn,\Brod].
Thus, we replace the functions $\tilde f_b (x_1, x_2,
\ldots, x_b)$ by
$$ g_b (n_1, n_2, \ldots, n_b)= (K)^{(1-b)/2}
\tilde f_b (n_1/K, n_2/K, \ldots, n_b/K)
$$
and
$$\int_0^1 dx \to {1\over K}\sum_{n=1}^K\,\,.
$$
According to our discussion of the boundary conditions,
we assume that $g_b$ is non-vanishing
only if all its arguments are positive integers ($n=0$ is excluded).
The cut-off eigenvalue problem assumes the form
$$\eqalign{& {2P^+P^-\over\mu} g_b (n_1, n_2, \ldots, n_b)= K
g_b (n_1, n_2, \ldots, n_b)\sum_{i=1}^b {1\over n_i}\cr
& -K y\biggl\{
\sum_{n'_1=1}^{n_1-1} {g_{b+1} (n'_1, n_1-n'_1, n_2, \ldots, n_b)\over
\sqrt {n_1 n'_1 (n_1-n'_1)}}+
\sum_{n'_2=1}^{n_2-1} {g_{b+1} (n_1, n'_2, n_2-n'_2, n_3, \ldots, n_b)\over
\sqrt {n_2 n'_2 (n_2-n'_2)}}+ \ldots \cr
&+ {g_{b-1} (n_1+n_2, n_3, \ldots, n_b)\over \sqrt {n_1 n_2 (n_1+n_2)}}+
{g_{b-1} (n_1, n_2+n_3, \ldots, n_b)\over \sqrt {n_2 n_3 (n_2+n_3)}}+
\ldots \biggr\}\,\,,
\cr }\eqn\coS$$
where $y=\lambda/(2\sqrt{\pi}\mu)$ is the dimensionless coupling.
The normalization condition becomes
$$\sum_{b=1}^\infty  b \sum_{n_1=1}^{K-1}\ldots \sum_{n_{b-1}=1}^{K-1}
|g_b (n_1, n_2, \ldots, n_{b-1}, K- \sum_{i=1}^{b-1} n_i)|^2 =1
\ .\eqn\conorm$$
The cut-off eigenvalue problem reduces to matrix diagonalization because
the $g$'s contain only a finite number of degrees of freedom, which
is equal to the number of partitions of $K$ into positive integers,
modulo cyclic permutations. All such partitions can be constructed
recursively with the help of a computer. We find that
for $K=10, 11, 12, 13, 14, 15, 16, 17, 18$ the number of degrees of freedom
is $107, 187, 351, 631, 1181, 2191, 4115, 7711, 14601$ respectively.
As could have been expected, it grows roughly exponentially with $K$.
The maximum number of bits is equal to
$K$. As the cut-off $K$ is taken to infinity, the possible values
of the longitudinal momentum fractions $x_i$ densely populate the
interval $(0, 1)$, which corresponds to the continuum limit,
\ie\ the eigenvalue problem of eq. \newlightS.

If we regard all the independent components of $g_b (n_1, n_2, \ldots,
n_b)$, with $b=1, 2, \ldots, K$, as components of a vector, then
the eigenvalue problem of eq. \coS\ is equivalent to diagonalizing
the matrix
$${2P^+P^-\over\mu} = K\left(V -y T\right) \ .
\eqn\newform$$
The matrix $V$ is diagonal, while $T$ is off-diagonal, connecting string
states whole length (number of bits) differs by 1. The entries of
$V$ and $T$ can be easily read off from eqs. \coS\ and \conorm.
Alternatively, as shown in ref. \DK, $V$, $T$, and a basis of normalized
states can be constructed in terms of creation and annihilation
operators.

\bigskip
\centerline{\caps 4. Numerical Work}
\bigskip

The objective of our numerical approach is to compute the low-lying
spectrum of the matrix \newform. First we fix the parameter $y$,
and vary the cut-off $K$ from 9 to 15, which is the highest value
accessible to us at present. Then, for each $y$, an extrapolation of the data
to $K\to\infty$ gives us an estimate of the spectrum of the
continuum eigenvalue problem of eq. \newlightS. Finally we plot the
estimated continuum eigenvalues versus $y$ looking for critical
behavior at some $y_c$.

What kind of singularity can we expect? Based on our experience with
the $c=1$ matrix model, we might expect that the spectrum becomes
dense at $y=y_c$, indicating the appearance of continuous Liouville
momentum $p_\phi$. The resulting string theory would be 3-dimensional,
with its spectrum given by
$${2P^+P^-\over {\cal T}}= p_\phi^2 + 4r -{1\over 6}
\eqn\guess$$
where $r$ runs over non-negative integers, and the tachyonic ground
state energy is found from the standard formula $(2-D)/6$.
${\cal T}$ is the string tension in the embedding dimensions, and $\mu$ can
be thought of as the ``bare'' string tension. It is possible that
${\cal T}$ exhibits some dependence on $y$, and it can even become singular
as $y\to y_c$. Depending on the behavior of ${\cal T}$ near the critical
point, there are two very different possibilities:

1) The string tension ${\cal T}$ is finite at the critical point $y_c$.
Then eq. \guess\ suggests that the spectrum of
eq. \newlightS\ should become continuous at $y=y_c$, starting at a tachyonic
value. This is the possibility suggested in ref. \DK.

2) ${\cal T}$ diverges at the critical point, so that all the low-lying
eigenvalues of eq. \newlightS\ tend to $-\infty$. This possibility
was not discussed in ref. \DK, but the divergence of
${\cal T}$ has in fact been advocated in earlier literature [\dur].

Now we present our current numerical results. Unfortunately, the
nature of the numerical problem makes it hard to decide with certainty
whether 1) or 2) is correct, unless values of $K$ considerably higher
than 15 can be reached. This is impossible with our present means, but
may be feasible on a supercomputer. Judging by the available data,
2) is the more likely possibility.

In fig. 1 we plot the lowest three eigenvalues of eq.
\coS\ versus $K$ for $y=0.2$.
Here the convergence to the continuum limit is found to be quite fast.
In general, we expect the dependence on $K$ to be of the
form
$$ m_i^2 (y, K) = m_i^2 (y, \infty) +\sum_{n=1}^\infty c_{i, n}(y) K^{-n}
$$
In order to estimate the continuum limit, we fitted the dependence
on $K$ at fixed $y$ to a ratio of two polynomials in $1/K$.
This fit allows us to extrapolate fig. 1 to arbitrarily large values
of $K$. This extrapolation saturates rapidly, showing intuitively expected
behavior.

In fig. 2 we plot the lowest eigenvalue,
$m_1^2/\mu$, versus $K$ for $y=0.2,~ 0.52,~ 0.65$.
These plots show the dependence of the rate of convergence on $y$.
For $y=0.52$ the convergence is much slower than for $y=0.2$.
For $y=0.65$, instead of being concave, the graph becomes convex,
so that the points no longer appear to converge to a finite limit as
$K\to\infty$.

In fig. 3 we show extrapolations of the lowest
eigenvalue to large $K$. For $y=0.50$ (fig. 3a) the extrapolation converges
to a value which is far below that found for $K=15$.
In fact, the continuum value is tachyonic, as
anticipated from a $D=3$ string theory.
For $y=0.55$ (fig. 3b) the extrapolation fails to converge.
It follows that
there should be a critical value of $y$, which lies between $0.50$
and $0.55$, where the extrapolated value of $m_1^2/\mu$ begins to
diverge towards $-\infty$. This is indeed what
happens.

In fig. 4 we plot the lowest three eigenvalues, extrapolated to infinite
$K$, versus $y$. While for $y=0$ the second and third eigenvalues are
degenerate, for $y>0$ they are separated by a finite gap, which is at
first too small to be visible on the figure.
The discreteness of the spectrum for $y>0$
indicates that the quanta are indeed bound into strings.
The lowest eigenvalue, $m_1^2/\mu$, begins to dip rapidly
for $y$ beyond $0.45$. First it becomes tachyonic, and then
diverges as $y$ approaches 0.53.
If we identify this value of $y$ with $y_c$, then the divergence of the
lowest eigenvalue seems consistent with the divergence of the string
tension suggested in the possibility 2).
Of course, it may be that the gaps in the
spectrum, $m_2^2-m_1^2$, $m_3^2-m_2^2$, etc. vanish before the
eigenvalues themselves diverge, but fig. 4 indicates
otherwise. We find that the extrapolated $m_2^2/\mu$ plotted vs. $y$
exhibits behavior similar to $m_1^2/\mu$, but diverges at a higher value of
$y$, and the gap $m_2^2-m_1^2$ does not vanish anywhere.
If possibility 2) is correct,
then all the eigenvalues should diverge at the same critical
value of $y$. The fact that this is not the case in fig. 4 may
be attributed to the lack of precision of our extrapolation.
We find, however, that the places where $m_1^2$ and $m_2^2$ diverge
become closer as the extrapolation includes the data for higher $K$.
In general, we have to add a cautionary note that
the precise shape of fig. 4 is not reliable. As $y$ approaches
$y_c$, the rate of convergence towards the continuum limit becomes
slower and slower, so that higher values of $K$ are needed for
a reliable estimate. Further numerical work is crucial
for testing our conjectures. We hope, however, that our numerical studies
provide a clue about the universal features of the continuum limit.

\bigskip
\centerline{\caps 5. Discussion}
\bigskip

The possibility 2) suggested in the previous section and
supported, to some extent, by the available data can be interpreted
as an infinite multiplicative renormalization of the string tension.
In other words, the effective string tension at the critical point
$y=y_c$ is infinitely larger than the bare string tension $\mu$.
A very similar scenario was advocated in ref. \dur\ and interpreted
as the branched polymer phase of random surfaces. Our results are
different, though, in that the spectrum at the critical point is
tachyonic, as expected from the continuum reasoning in
eq. \guess.

The divergence of the string tension at $y=y_c$ could probably
be cured by appropriately scaling the bare string tension $\mu$
to zero. However, the resulting string theory with finite spectrum
would still be tachyonic.

Another problem, which is perhaps even more severe, comes from the
states that transform non-trivially under $SU(N)$, such as the
adjoint representation states of the form
$$\sum_{b=1}^\infty \int_0^{P^+} dk_1 \ldots dk_b
\delta\left (\sum_{i=1}^b k_i-P^+\right ) f_b (k_1, \ldots, k_b)
 N^{(1-b)/2} a^\dagger_{ij}(k_1)
a^\dagger_{jk}(k_2) \ldots
a^\dagger_{rs}(k_{b-1})
a^\dagger_{st}(k_{b}) |0 \rangle
\ . \eqn\eq$$
As discussed previously, the $SU(N)$ \ns\ states
cannot be identified with closed strings. One indication of that
is their diverging degeneracy factors in the $N\to\infty$ limit.
Perhaps the \ns s can be thought of as  closed strings that
have disintegrated into separate bits. Of course, one way of preventing
this is through a confinement phenomenon which would push their
energies to infinity in the continuum limit. In fact, confinement
is at work in the $c=1$ string theory, where the \ns\ energies diverge
logarithmically in the cut-off [\Vort, \bk].

Could the $c=2$ model of eq. \action\ also expel the \ns\ states to
infinite energy? The answer is negative, as can be seen from
a simple variational calculation. Taking $a_{ij}^\dagger (P^+)|0 \rangle$
as the variational state, and evaluating the matrix element of
$2P^+ P^-/\mu$,
gives an upper bound of 1 for the
lowest eigenvalue.
Therefore, there is no confinement, and the \ns\ states
probably cause additional problems for the interacting string
theory. Some heuristic arguments with similar conclusions
were made in the context of $c>1$ models with discrete target
spaces [\Vort]. The proliferation of the \ns s was associated with
the \KT\ vortices which wind around the target space plaquettes
of lattice size. An advantage of the \lc\ approach is that
the \ns\ spectrum can be found with the same technique as
the singlet spectrum. It would be quite useful to carry out
a more detailed numerical study of the spectrum in the adjoint
representation.

The lack of confinement of the \ns\ states can be traced
to the fact that the action \action\ has only the global
$SU(N)$ symmetry, which is not gauged.
Gauging the $SU(N)$ leads to the confinement,
and the non-singlets are pushed to infinite energy [\Gauged].
The effect of gauging the $SU(N)$ symmetry on the \td\ theory of
eq. \action\ is  to give the action
$$ S_{gauged}=\int d^2 x \Tr \left ({1\over 4g^2}F_{\alpha\beta}^2+
\half (\partial_\alpha M+i[A_\alpha, M])^2+\half \mu M^2-
{1\over 3}{\lambda\over\sqrt N} M^3 \right )\ .
\eqn\nac$$
Light-cone quantization of this action was investigated in ref.
\Gauged.
There a linear \lc\ Schroedinger equation for the $N\to\infty$ limit,
similar to eq. \lightS, was derived. At the moment we see no
major obstacles to studying this equation using the
discretized longitudinal momentum cut-off. Perhaps, the theory \nac\
has new types of critical behavior which lead to more interesting
string theories than those originating from the model \action.

\ack
I. R. K. wishes to thank all the organizers of the
7th Nishinomiya-Yukawa Memorial Symposium, and especially
M. Ninomiya, for their warm hospitality and for creating
a stimulating scientific environment. We are indebted to
S. Dalley, B. Durhuus and A. Polyakov for discussions, and to G. Bhanot for his
advice on computing.
The authors are supported in part by
NSF Presidential Young Investigator Award PHY-9157482 and
James S. McDonnell Foundation grant No. 91-48. I. R. K. is also supported by
an A. P. Sloan Foundation Research Fellowship
and DOE grant DE-AC02-76WRO3072,

\singlespace
\refout

\bigskip
\centerline{FIGURE CAPTIONS}
\bigskip

1. The lowest three eigenvalues of eq.  \coS\ versus $K$ for $y=0.2$.

2. The plots of the lowest eigenvalue of eq. \coS\ versus $K$
for $y=0.2,~ 0.52,~ 0.65$.

3. Extrapolation to large $K$ of the lowest eigenvalue for
a) $y=0.50$; b) $y=0.55$.

4. The lowest three eigenvalues, extrapolated to infinite $K$,
versus $y$.
\bye